\documentclass[useAMS,usenatbib,usegraphicx]{mn2e}

\usepackage{mathptmx,subfigure,jab}
\usepackage[fleqn]{amsmath}

\title[Solar wind magnetic field rotations]{Magnetic field rotations in the solar wind at kinetic scales}
\author[C. H. K. Chen, L.~Matteini, D.~Burgess and T. S. Horbury]{C. H. K. Chen,$^{1}$\thanks{E-mail: christopher.chen@imperial.ac.uk} L.~Matteini,$^{1}$ D.~Burgess$^{2}$ and T. S. Horbury$^{1}$\\
$^{1}$Department of Physics, Imperial College London, London SW7 2AZ, UK\\
$^{2}$Astronomy Unit, Queen Mary University of London, London E1 4NS, UK}

\begin{document}
\date{}
\pagerange{}
\pubyear{2015}
\maketitle

\begin{abstract}
The solar wind magnetic field contains rotations at a broad range of scales, which have been extensively studied in the MHD range. Here we present an extension of this analysis to the range between ion and electron kinetic scales. The distribution of rotation angles was found to be approximately lognormal, shifting to smaller angles at smaller scales almost self-similarly, but with small, statistically significant changes of shape. The fraction of energy in fluctuations with angles larger than $\alpha$ was found to drop approximately exponentially with $\alpha$, with e-folding angle $9.8^\circ$ at ion scales and $0.66^\circ$ at electron scales, showing that large angles ($\alpha > 30^\circ$) do not contain a significant amount of energy at kinetic scales. Implications for kinetic turbulence theory and the dissipation of solar wind turbulence are discussed.
\end{abstract}

\section{Introduction}

The solar wind magnetic field displays broadband fluctuations that are in many ways consistent with our current understanding of plasma turbulence \citep[e.g.,][]{horbury12,bruno13,alexandrova13a}. It also contains many small-scale structures, which may or may not arise from the turbulent dynamics. Determining the properties of these kinetic scale fluctuations is important for understanding the turbulent cascade and how its energy is dissipated, as well as the general structure of the solar wind.

Large changes in the solar wind magnetic field direction are sometimes called ``directional discontinuities'' \citep{burlaga69} and are frequently \citep[e.g.,][]{burlaga71,neugebauer84,tsurutani96,horbury01b,paschmann13} classified as rotational discontinuities (RDs) or tangential discontinuities (TDs) depending on which of these discontinuous solutions of ideal magnetohydrodynamics (MHD) \citep{landau60} they more resemble. Their origin is debated \citep{burlaga69,burlaga71,bruno01,vasquez07b,borovsky08,greco08a,greco09a,owens11,zhdankin12a,zhdankin12b,borovsky12b,malaspina12,arnold13}, in particular, whether they are generated in situ or represent plasma boundaries arising from processes at the Sun. For example, it has been proposed \citep[e.g.,][]{bruno01,borovsky08} that large angle changes represent flux tube boundaries in a filamentary picture of the solar wind originating from early solar energetic particle observations \citep{bartley66,mccracken66}. However, it has also been suggested that their waiting times and angular distributions are consistent with MHD turbulence  \citep{vasquez07b,greco08a,greco09a,zhdankin12a,zhdankin12b}, suggesting that the majority are generated in situ. A recent simulation found MHD turbulence to produce both RDs and TDs, although RDs were found to be more numerous \citep{zhang15}.

It is perhaps more instructive to examine the full distribution of magnetic field rotation angles, rather than just the large ones. \citet{borovsky08} fitted the full distribution to a double exponential, interpreting the one dominating at small angles to be due to turbulence and the one at large angles to be due to flux tube boundaries. \citet{borovsky12b}, however, later showed that plasma boundaries contribute only a small fraction of the distribution at all angles. \citet{zhdankin12a} showed that rather than an exponential, MHD turbulence simulations produce an angle distribution similar to the full distribution in the solar wind. \citet{zhdankin12b} then showed that the solar wind distribution could be fit to a single lognormal population, and moreover that the shape of this is independent of scale over the three decades of the MHD inertial range, suggesting an underlying universal description.

One of the questions addressed in this Letter is how the rotation angle distribution continues to below ion kinetic scales. Here the spectra of magnetic and density fluctuations steepen \citep[e.g.,][]{coleman68,russell72,leamon98a,smith06a,chen13a,safrankova13a,bruno14a,bruno14b,chen14b} and form a power law range with spectral index $\approx -2.8$ \citep{denskat83,kiyani09a,chen10b,alexandrova12,chen12a,sahraoui13a,chen13c}. While this is generally consistent with scaling theories of strong sub-ion scale turbulence \citep{vainshtein73,biskamp96,cho04,schekochihin09,chen10a,boldyrev12b,boldyrev13a}, this range is less well understood than the turbulence at MHD scales. At kinetic scales the shapes of the distributions of magnetic field component and density fluctuations do not vary much with scale \citep{kiyani09a,wu13,chen14a}, but angle distributions have not previously been measured.

A related question is what type of structures the large angle changes correspond to. MHD turbulence is thought to generate sheet like structures, and both statistical \citep[e.g.,][]{biskamp03} and dynamical \citep{howes15} models have proposed to explain this. Recent work has shown that kinetic scale turbulence may do the same, for example, 3D fluid \citep{boldyrev12b} and kinetic \citep{tenbarge13a} simulations show the development of 2D sheets at electron scales, although 1D filaments have also been reported \citep{meyrand13}. It has been suggested that some of the current sheets formed in MHD turbulence \citep[e.g.,][]{carbone90,cowley97,servidio09} and sub-ion scale turbulence \citep{haynes14} may be undergoing reconnection. Evidence of small scale reconnection has been reported in the magnetosheath \citep{retino07,sundkvist07} and solar wind \citep{gosling12,perri12a,osman14b,xu15b}, possibly as part of the turbulent cascade. It has been suggested that a significant amount \citep{leamon00} or even the majority \citep{matthaeus03a} of energy in MHD turbulence is dissipated via reconnection. However, others have suggested that enhanced dissipation can occur at current sheets through other processes such as Landau damping \citep{tenbarge13a}.

In this Letter, we extend the analysis of the distribution of magnetic field rotations into the kinetic range between ion and electron scales. The scale-dependence of the distribution is examined, as well as the fraction of energy contained at the different angles. The implications for dissipation of turbulence and reconnection are also discussed.

\section{Data Set}

The analysis was performed on a 7 hour interval of data from the \emph{Cluster} spacecraft \citep{escoubet01} starting at 2003-02-12 23:00:00, when the four spacecraft were in the free solar wind (out of Earth's foreshock). The typical spacecraft-frame correlation time at 1 AU is $\sim$ 30 mins \cite[e.g.,][]{osman14b} so the interval covers many turbulence correlation lengths. For the magnetic field $\mathbf{B}$, data from the DC magnetometers (FGM) \citep{balogh01} and AC magnetometers (STAFF) \citep{cornilleau-wehrlin03} were combined into a single data set of 0.04 s resolution for each spacecraft, following the procedure given in \citet{chen10b}. The spectrum of magnetic fluctuations remains at least an order of magnitude above the nominal instrumental noise floors over the entire range.

To determine the plasma microscales, additional particle moment data is required. The electron density $n_\mathrm{e}$ was obtained from the plasma frequency determined from the high frequency electric field spectrum analyzer (WHISPER) \citep{decreau01} and the perpendicular electron temperature $T_\mathrm{\perp,e}$ from the electron electrostatic analysers (PEACE) \citep{johnstone97}. Since data from the appropriate ion electrostatic analysers were not available, the ion (proton) velocity $\mathbf{v}_\mathrm{i}$ and perpendicular temperature $T_\mathrm{\perp,i}$ were obtained from the Faraday cups (SWE) \citep{ogilvie95} on the upstream \emph{Wind} spacecraft \citep{acuna95}, shifted in time so that magnetic field features measured by both spacecraft were aligned. The mean parameters for this interval are: $|\mathbf{B}|=10$ nT, $n_\mathrm{e}=12$ cm$^{-3}$, $|\mathbf{v}_\mathrm{i}|=390$ km s$^{-1}$, $T_\mathrm{\perp,i}=5.6$ eV, $T_\mathrm{\perp,e}=13$ eV; the total perpendicular plasma beta is $\beta_\perp=0.88$. 

\section{Results}

\begin{figure}
\includegraphics[width=\columnwidth,trim=0 0 0 0,clip]{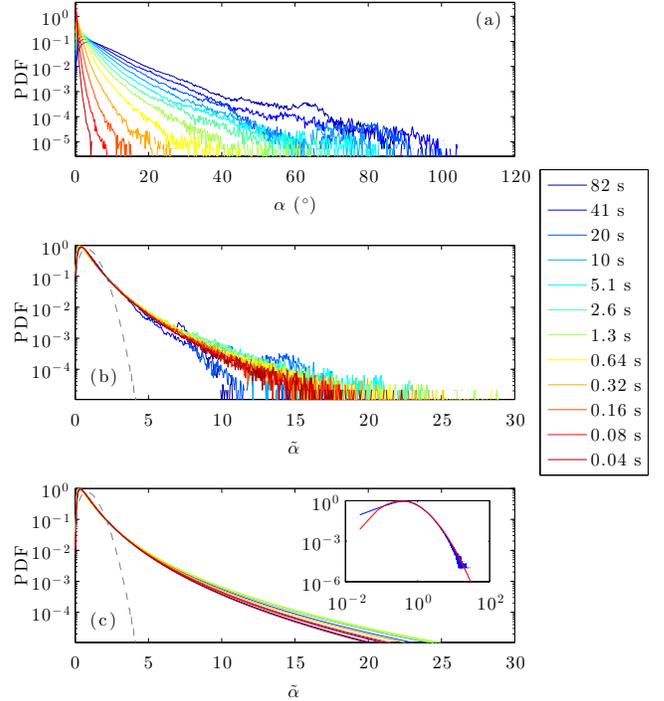}
\caption{\label{fig:alpha}(a) PDFs of magnetic field rotation angle $\alpha$ for different time lags $\tau$. (b) PDFs of $\tilde{\alpha}=\alpha/\left<\alpha\right>$. (c) Lognormal fits of PDFs of $\tilde{\alpha}$; the inset shows the measured PDF (blue) and fit (red) for $\tau=0.04$ s. The distribution for a Gaussian process is also shown (grey dashed).}
\end{figure}

The magnetic field rotation angle over time scale $\tau$ and at time $t$ is
\begin{equation}
\alpha(t,\tau)=\cos^{-1}\left[\frac{\mathbf{B}(t)\cdot\mathbf{B}(t+\tau)}{|\mathbf{B}(t)||\mathbf{B}(t+\tau)|}\right].
\end{equation}
Fig.~\ref{fig:alpha}a shows the probability density functions (PDFs) of $\alpha$ at scales $\tau$ from 0.04 s to 82 s. Data from spacecraft 2--4 are included in the PDFs (spacecraft 1 data have significant spin tone). The distribution moves to smaller angles at smaller scales, as expected, since the fluctuation amplitudes decrease.

Under the \citet{taylor38} hypothesis, spacecraft-frame time scales correspond to spatial scales in the plasma. While this may break down if the turbulence evolves quickly as it passes the spacecraft, it is thought to be a good approximation for anisotropic kinetic Alfv\'en turbulence \citep{howes14b,klein14b}, which measurements suggest is the dominant component at kinetic scales \citep{chen10b,chen13c,telloni15}. Under this approximation, $\tau\sim 1.3$~s corresponds to ion scales (e.g., gyroradius, inertial length) and  $\tau\sim 0.04$~s to electron scales. The data in Fig.~\ref{fig:alpha}a show that the maximum value of $\alpha$ in this interval is 65$^\circ$ at ion scales and 5.4$^\circ$ at electron scales; the mean values are 2.2$^\circ$ and 0.17$^\circ$ respectively.

To test the self-similarity of $\alpha$ over the kinetic range, Fig.~\ref{fig:alpha}b shows the distributions of $\tilde{\alpha}=\alpha/\left<\alpha\right>$, i.e., $\alpha$ normalised to its mean value, where the angular brackets denote an average over $t$. Apart from variations at large angles for large $\tau$, the distributions appear to be quite self-similar (although non-Gaussian) with only small changes with scale. However, a quantitative test is required to assess the degree of self-similarity. The kurtosis of the $\tilde{\alpha}$ distribution was measured directly, although this was found to be unreliable due to the finite interval length \citep{horbury97a,dudokdewit04}. Therefore, the distributions were fit to various functions. The best fit was obtained for a lognormal distribution,
\begin{equation}
f(x)=\frac{1}{x\sigma\sqrt{2\pi}}\exp\left[-\frac{(\ln x-\mu)^2}{2\sigma^2}\right],
\end{equation}
similar to previous findings for the MHD range \citep[e.g.,][]{bruno04,zhdankin12b}. The fitted distributions are shown in Fig.~\ref{fig:alpha}c; they are able to capture most of the features of the measurements, although the smallest angles are slightly underestimated (see inset). The fit is better than a double exponential \citep[used by][]{borovsky08}, and has fewer free parameters.

\begin{figure}
\includegraphics[width=\columnwidth,trim=0 0 0 0,clip]{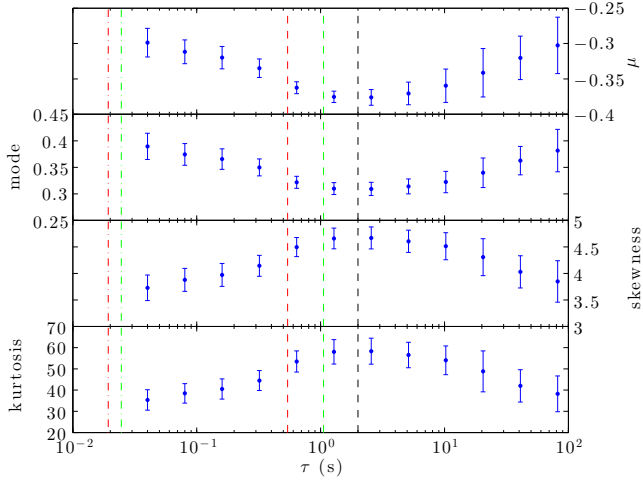}
\caption{\label{fig:fits}Lognormal fit parameter $\mu$ and other inferred parameters of the distribution of $\tilde{\alpha}$ as a function of time lag $\tau$. Scales corresponding to the proton (dashed) and electron (dash-dotted) gyroradii (red) and inertial lengths (green) are marked, along with the measured spectral break (black dashed).}
\end{figure}

Fig.~\ref{fig:fits} shows the fit results as a function of $\tau$. Since at all scales the mean of $\tilde{\alpha}$ is 1, the fit effectively has one free parameter, so only $\mu$ is shown. The mode, skewness and kurtosis calculated from the fit parameters are also shown. The error bars represent the standard deviations calculated from 7 non-overlapping 1-hour sub-intervals. It can be seen that there is a small but statistically significant variation in the parameters with scale. For example, the kurtosis increases from large to small scales, peaks at ion scales, then decreases towards electron scales; a behaviour consistent with the kurtosis of the magnetic field component fluctuations in this interval (not shown here) and other works \citep[e.g.,][]{wu13}.

\begin{figure}
\includegraphics[width=\columnwidth,trim=0 0 0 0,clip]{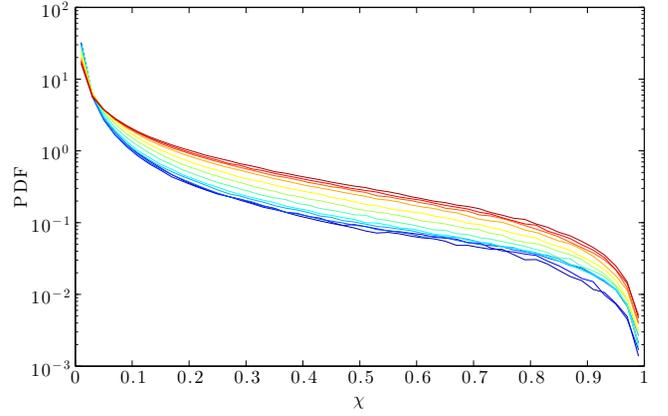}
\caption{\label{fig:chi}PDFs of $\chi$ (Equation \ref{eq:chi}) at different time lags $\tau$. Colors are the same as those in Fig.~\ref{fig:alpha}.}
\end{figure}

\citet{zhdankin12b} used the parameter 
\begin{equation}
\label{eq:chi}
\chi(t,\tau)=\frac{|\delta B/B-2\sin(\alpha/2)|}{\delta B/B},
\end{equation}
where $\delta B/B=|\mathbf{B}(t+\tau)-\mathbf{B}(t)|/|\mathbf{B}(t)|$, to measure the degree to which the magnetic field changes magnitude during a rotation. $\chi=0$ corresponds to a pure rotation and $0<\chi\leq 1$ corresponds to a rotation with change in field strength (note that this does not distinguish TDs from RDs). Fig.~\ref{fig:chi} shows the PDFs of $\chi$ over the same range of scales as Figures \ref{fig:alpha} and \ref{fig:fits}. At all scales the distribution peaks at $\chi =0$ showing that the fluctuations are predominantly pure rotations. However, the peak becomes less strong towards smaller scales, consistent with the turbulence being more compressive in the kinetic range. Since the fluctuations are mostly pure rotations, $\delta B/B\approx\alpha$ for small $\alpha$, and indeed the results for the distributions of $\delta B/B$ match the above results for the distributions of $\alpha$.

Finally, to help understand the distribution of energy and how it is dissipated at kinetic scales, the fluctuation energy contained in the magnetic rotations was measured. Fig.~\ref{fig:fraction} shows the fraction of angles larger than $\alpha$, and the fraction of magnetic fluctuation energy $|\mathbf{B}(t+\tau)-\mathbf{B}(t)|^2$ that they contain, at ion and electron scales. For both scales, the energy drops approximately exponentially with $\alpha$. A similar result was found for the energy dissipation in current structures in a recent 3D RMHD simulation \citep{zhdankin14}, although the comparison is less good with a 2D PIC simulation \citep{wan12a}, perhaps suggesting that 3D simulations are better able to capture the properties of turbulent structures formed in the solar wind. The e-folding value is 9.8$^\circ$ for ion scales and 0.66$^\circ$ for electron scales, showing that throughout the kinetic range there is only a small amount of energy in large angles ($\alpha > 30^\circ$), although since the typical angles are small, those a few times larger than the mean still contain a significant fraction of the energy.

The above analysis was performed on several other intervals of data and found to be consistent with that presented here.

\section{Discussion}

We have shown that the distribution of magnetic field rotations is approximately lognormal from ion to electron scales, however there is a small, but statistically significant, change in shape with scale, e.g., the kurtosis decreases by a factor of 1.6 from ion to electron scales, consistent with the magnetic field components \citep[e.g.,][]{wu13}. The interval is not long enough to probe the full MHD range to enable a full comparison with the results of \citet{zhdankin12b}, although the lognormal fit parameters are somewhat different in the range where the analyses overlap (3~s to 82~s). This may be related to the fact that the current study is based on 7 hours of data (a single ``stream''), while the study of \citet{zhdankin12b} used 6 years, meaning that the distributions there may in part reflect the varying driving conditions rather than being solely an effect of the turbulent cascade.

\begin{figure}
\includegraphics[width=\columnwidth,trim=0 0 0 0,clip]{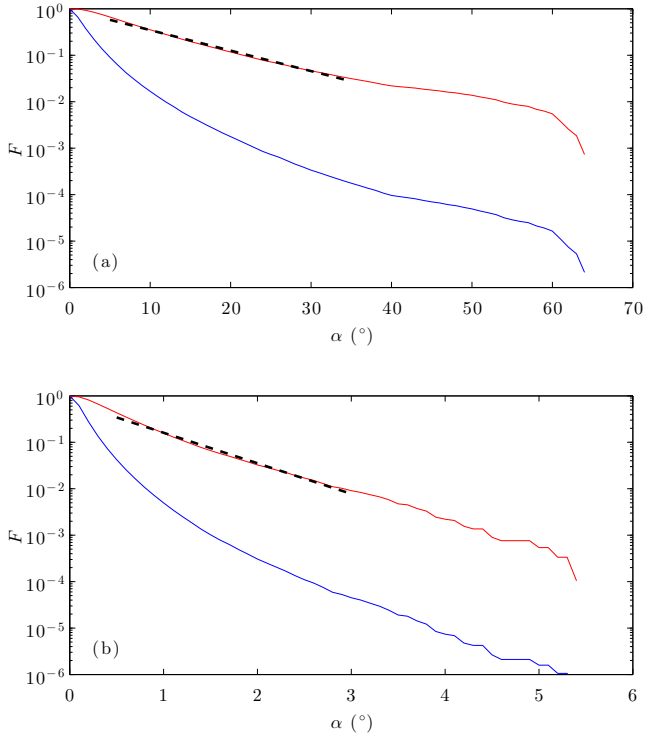}
\caption{\label{fig:fraction}(a) Fraction $F$ of ion scale fluctuations with an angle $>\alpha$ (blue) and fraction of magnetic fluctuation energy in those fluctuations (red). The black dashed line shows an exponential fit $F=0.96\exp (-\alpha/9.8)$. (b) Same for electron scales with fit $F=0.73\exp (-\alpha/0.66)$.}
\end{figure}

Figures \ref{fig:alpha} and \ref{fig:fraction} also show that the rotation angles at kinetic scales are relatively small, much smaller than those obtained by \citet{perri12a}. The reason for this is that \citet{perri12a} used only data from STAFF in their calculation of $\alpha$ (rather than the combined FGM-STAFF data set used here), which effectively provides a magnetic field time series high pass filtered at $\sim$ 1 Hz. Due to the lack of DC field, the values of $\alpha$ obtained by \citet{perri12a} are much larger than the true rotation angles, as shown here. The larger fluctuations of the distribution may still correspond to particular types of structures, although current measurements are not sufficient to unambiguously determine their geometry.

Given the small amount of energy in large angle fluctuations at kinetic scales (Fig.~\ref{fig:fraction}), it might be questioned how significant reconnection is for dissipating turbulence in the solar wind. While this cannot be answered with the current data set, knowing the distribution of angles can help constrain this possibility. For example, reconnection, in the traditional sense, is known to be suppressed at small angles if $\beta$ fluctuations across the current sheet are large enough \citep{phan10} due to the diamagnetic drift of the x-line being faster than the outflow speed \citep{swisdak03,swisdak10}. High resolution, low noise particle data are required to check this condition for turbulent fluctuations at kinetic scales.

Reconnection has been identified in the solar wind from Alfv\'enic jets associated with bifurcated magnetic field rotations \citep[e.g.,][]{gosling12}, although typically only a few events per day are found overall \citep[e.g.,][]{gosling07b,osman14b}, with fewer than one per day in the purer fast wind \citep{gosling07a}. While reconnection at small angles is occasionally detected \citep{gosling13}, this is only for very low $\beta$ plasma not typical of the solar wind at 1 AU. The typical separation between measured reconnection events is one to two orders of magnitude larger than the turbulence correlation length and such rarity led \citet{gosling07a} to question whether it can be a major way for the turbulence to be dissipated. High resolution magnetic field data have shown that only a small number of reconnection events (identified from bifurcated rotations) are missed due to limited resolution particle data \citep{gosling08}. The data presented here were examined for bifurcated magnetic rotations, although no clear signatures were found. It may be the case that the amount of turbulent energy dissipated via reconnection (rather than dissipated directly) depends on parameters of the system, such as the length of the inertial range or plasma $\beta$. However, even if some energy conversion occurs through reconnection, how it is finally dissipated still remains to be identified \citep[e.g,][]{loureiro13,haynes14,drake14}.

Finally, knowing the full distributions of magnetic fluctuations at kinetic scales allows one of the important assumptions of several approaches to solar wind turbulence to be tested. It is sometimes assumed, for example in reduced non-linear models \citep[e.g.,][]{howes06,schekochihin09,boldyrev13a,kunz15} and linear theory \citep[e.g.,][]{gary09}, that $\delta B/B\ll 1$. While this is well satisfied for the typical fluctuation (on average $\delta B/B=0.040$ at ion scales and $\delta B/B=0.0034$ at electron scales for this interval), it has been questioned to what extent this is valid for the larger amplitude fluctuations which are an important part of the cascade. In the current interval, at ion scales 1.1\% of the fluctuations, containing 30\% of the energy, have $\delta B/B>0.2$ and at electron scales all fluctuations satisfy $\delta B/B<0.1$. Therefore, one could conclude that the assumption becomes valid for the majority of fluctuations around ion scales, and is very well satisfied by electron scales.

\section*{acknowledgments}
CHKC is supported by an Imperial College Junior Research Fellowship. We acknowledge support provided to ISSI/ISSI-BJ Team 304. \emph{Cluster} data were obtained from the Cluster Active Archive ({http://caa.estec.esa.int}). We thank S. Boldyrev and ISSI Team 304 for useful discussions.

\bibliographystyle{mn2e}
\bibliography{bibliography}

\begin{thebibliography}{93}
\expandafter\ifx\csname natexlab\endcsname\relax\def\natexlab#1{#1}\fi

\bibitem[{{Acu{\~n}a} {et~al}\mbox{.}(1995){Acu{\~n}a}, {Ogilvie}, {Baker},
  {Curtis}, {Fairfield}, \& {Mish}}]{acuna95}
{Acu{\~n}a} M.~H., {Ogilvie} K.~W., {Baker} D.~N., {Curtis} S.~A., {Fairfield}
  D.~H., {Mish} W.~H., 1995, \ssr, 71, 5

\bibitem[{{Alexandrova} {et~al}\mbox{.}(2013){Alexandrova}, {Chen},
  {Sorriso-Valvo}, {Horbury}, \& {Bale}}]{alexandrova13a}
{Alexandrova} O., {Chen} C.~H.~K., {Sorriso-Valvo} L., {Horbury} T.~S., {Bale}
  S.~D., 2013, \ssr, 178, 101

\bibitem[{{Alexandrova} {et~al}\mbox{.}(2012){Alexandrova}, {Lacombe},
  {Mangeney}, {Grappin}, \& {Maksimovic}}]{alexandrova12}
{Alexandrova} O., {Lacombe} C., {Mangeney} A., {Grappin} R., {Maksimovic} M.,
  2012, \apj, 760, 121

\bibitem[{{Arnold} {et~al}\mbox{.}(2013){Arnold}, {Li}, {Li}, \&
  {Yan}}]{arnold13}
{Arnold} L., {Li} G., {Li} X., {Yan} Y., 2013, \apj, 766, 2

\bibitem[{{Balogh} {et~al}\mbox{.}(2001){Balogh}, {Carr}, {Acu{\~n}a},
  {Dunlop}, {Beek}, {Brown}, {Forna{\c c}on}, {Georgescu}, {Glassmeier},
  {Harris}, {Musmann}, {Oddy}, \& {Schwingenschuh}}]{balogh01}
{Balogh} A. {et~al.}, 2001, \ang, 19, 1207

\bibitem[{{Bartley} {et~al}\mbox{.}(1966){Bartley}, {Bukata}, {McCracken}, \&
  {Rao}}]{bartley66}
{Bartley} W.~C., {Bukata} R.~P., {McCracken} K.~G., {Rao} U.~R., 1966, \jgr,
  71, 3297

\bibitem[{{Biskamp}(2003)}]{biskamp03}
{Biskamp} D., 2003, {Magnetohydrodynamic Turbulence}. Cambridge University
  Press

\bibitem[{{Biskamp}, {Schwarz} \& {Drake}(1996){Biskamp}, {Schwarz}, \&
  {Drake}}]{biskamp96}
{Biskamp} D., {Schwarz} E., {Drake} J.~F., 1996, \prl, 76, 1264

\bibitem[{{Boldyrev} {et~al}\mbox{.}(2013){Boldyrev}, {Horaites}, {Xia}, \&
  {Perez}}]{boldyrev13a}
{Boldyrev} S., {Horaites} K., {Xia} Q., {Perez} J.~C., 2013, \apj, 777, 41

\bibitem[{{Boldyrev} \& {Perez}(2012)}]{boldyrev12b}
{Boldyrev} S., {Perez} J.~C., 2012, \apjl, 758, L44

\bibitem[{{Borovsky}(2008)}]{borovsky08}
{Borovsky} J.~E., 2008, \jgr, 113, A08110

\bibitem[{{Borovsky}(2012)}]{borovsky12b}
{Borovsky} J.~E., 2012, \jgr, 117, A06107

\bibitem[{{Bruno} \& {Carbone}(2013)}]{bruno13}
{Bruno} R., {Carbone} V., 2013, \lrsp, 10, 2

\bibitem[{{Bruno} {et~al}\mbox{.}(2004){Bruno}, {Carbone}, {Primavera},
  {Malara}, {Sorriso-Valvo}, {Bavassano}, \& {Veltri}}]{bruno04}
{Bruno} R., {Carbone} V., {Primavera} L., {Malara} F., {Sorriso-Valvo} L.,
  {Bavassano} B., {Veltri} P., 2004, \ang, 22, 3751

\bibitem[{{Bruno} {et~al}\mbox{.}(2001){Bruno}, {Carbone}, {Veltri},
  {Pietropaolo}, \& {Bavassano}}]{bruno01}
{Bruno} R., {Carbone} V., {Veltri} P., {Pietropaolo} E., {Bavassano} B., 2001,
  \planss, 49, 1201

\bibitem[{{Bruno} \& {Trenchi}(2014)}]{bruno14a}
{Bruno} R., {Trenchi} L., 2014, \apjl, 787, L24

\bibitem[{{Bruno}, {Trenchi} \& {Telloni}(2014){Bruno}, {Trenchi}, \&
  {Telloni}}]{bruno14b}
{Bruno} R., {Trenchi} L., {Telloni} D., 2014, \apjl, 793, L15

\bibitem[{{Burlaga}(1969)}]{burlaga69}
{Burlaga} L.~F., 1969, \solphys, 7, 54

\bibitem[{{Burlaga}(1971)}]{burlaga71}
{Burlaga} L.~F., 1971, \jgr, 76, 4360

\bibitem[{{Carbone}, {Veltri} \& {Mangeney}(1990){Carbone}, {Veltri}, \&
  {Mangeney}}]{carbone90}
{Carbone} V., {Veltri} P., {Mangeney} A., 1990, \pof, 2, 1487

\bibitem[{{Chen} {et~al}\mbox{.}(2013{\natexlab{a}}){Chen}, {Boldyrev}, {Xia},
  \& {Perez}}]{chen13c}
{Chen} C.~H.~K., {Boldyrev} S., {Xia} Q., {Perez} J.~C., 2013{\natexlab{a}},
  \prl, 110, 225002

\bibitem[{{Chen} {et~al}\mbox{.}(2010{\natexlab{a}}){Chen}, {Horbury},
  {Schekochihin}, {Wicks}, {Alexandrova}, \& {Mitchell}}]{chen10b}
{Chen} C.~H.~K., {Horbury} T.~S., {Schekochihin} A.~A., {Wicks} R.~T.,
  {Alexandrova} O., {Mitchell} J., 2010{\natexlab{a}}, \prl, 104, 255002

\bibitem[{{Chen} {et~al}\mbox{.}(2013{\natexlab{b}}){Chen}, {Howes}, {Bonnell},
  {Mozer}, {Klein}, \& {Bale}}]{chen13a}
{Chen} C.~H.~K., {Howes} G.~G., {Bonnell} J.~W., {Mozer} F.~S., {Klein} K.~G.,
  {Bale} S.~D., 2013{\natexlab{b}}, \aipcp, 1539, 143

\bibitem[{{Chen} {et~al}\mbox{.}(2014{\natexlab{a}}){Chen}, {Leung},
  {Boldyrev}, {Maruca}, \& {Bale}}]{chen14b}
{Chen} C.~H.~K., {Leung} L., {Boldyrev} S., {Maruca} B.~A., {Bale} S.~D.,
  2014{\natexlab{a}}, \grl, 41, 8081

\bibitem[{{Chen} {et~al}\mbox{.}(2012){Chen}, {Salem}, {Bonnell}, {Mozer}, \&
  {Bale}}]{chen12a}
{Chen} C.~H.~K., {Salem} C.~S., {Bonnell} J.~W., {Mozer} F.~S., {Bale} S.~D.,
  2012, \prl, 109, 035001

\bibitem[{{Chen} {et~al}\mbox{.}(2014{\natexlab{b}}){Chen}, {Sorriso-Valvo},
  {{\v S}afr{\'a}nkov{\'a}}, \& {N{\v e}me{\v c}ek}}]{chen14a}
{Chen} C.~H.~K., {Sorriso-Valvo} L., {{\v S}afr{\'a}nkov{\'a}} J., {N{\v
  e}me{\v c}ek} Z., 2014{\natexlab{b}}, \apjl, 789, L8

\bibitem[{{Chen} {et~al}\mbox{.}(2010{\natexlab{b}}){Chen}, {Wicks}, {Horbury},
  \& {Schekochihin}}]{chen10a}
{Chen} C.~H.~K., {Wicks} R.~T., {Horbury} T.~S., {Schekochihin} A.~A.,
  2010{\natexlab{b}}, \apjl, 711, L79

\bibitem[{{Cho} \& {Lazarian}(2004)}]{cho04}
{Cho} J., {Lazarian} A., 2004, \apjl, 615, L41

\bibitem[{{Coleman}(1968)}]{coleman68}
{Coleman} P.~J., 1968, \apj, 153, 371

\bibitem[{{Cornilleau-Wehrlin} {et~al}\mbox{.}(2003){Cornilleau-Wehrlin},
  {Chanteur}, {Perraut}, {Rezeau}, {Robert}, {Roux}, {de Villedary}, {Canu},
  {Maksimovic}, {de Conchy}, {Lacombe}, {Lefeuvre}, {Parrot}, {Pin{\c c}on},
  {D{\'e}cr{\'e}au}, {Harvey}, {Louarn}, {Santolik}, {Alleyne}, {Roth},
  {Chust}, {Le Contel}, \& {STAFF Team}}]{cornilleau-wehrlin03}
{Cornilleau-Wehrlin} N. {et~al.}, 2003, \ang, 21, 437

\bibitem[{{Cowley}, {Longcope} \& {Sudan}(1997){Cowley}, {Longcope}, \&
  {Sudan}}]{cowley97}
{Cowley} S.~C., {Longcope} D.~W., {Sudan} R.~N., 1997, \physrep, 283, 227

\bibitem[{{D{\'e}cr{\'e}au} {et~al}\mbox{.}(2001){D{\'e}cr{\'e}au}, {Fergeau},
  {Krasnoselskikh}, {Le Guirriec}, {L{\'e}v{\^e}que}, {Martin},
  {Randriamboarison}, {Rauch}, {Sen{\'e}}, {S{\'e}ran}, {Trotignon}, {Canu},
  {Cornilleau}, {de F{\'e}raudy}, {Alleyne}, {Yearby}, {M{\"o}gensen},
  {Gustafsson}, {Andr{\'e}}, {Gurnett}, {Darrouzet}, {Lemaire}, {Harvey},
  {Travnicek}, \& {Whisper Experimenters}}]{decreau01}
{D{\'e}cr{\'e}au} P.~M.~E. {et~al.}, 2001, \ang, 19, 1241

\bibitem[{{Denskat}, {Beinroth} \& {Neubauer}(1983){Denskat}, {Beinroth}, \&
  {Neubauer}}]{denskat83}
{Denskat} K.~U., {Beinroth} H.~J., {Neubauer} F.~M., 1983, \jg, 54, 60

\bibitem[{{Drake} \& {Swisdak}(2014)}]{drake14}
{Drake} J.~F., {Swisdak} M., 2014, \pop, 21, 072903

\bibitem[{{Dudok de Wit}(2004)}]{dudokdewit04}
{Dudok de Wit} T., 2004, \pre, 70, 055302

\bibitem[{{Escoubet}, {Fehringer} \& {Goldstein}(2001){Escoubet}, {Fehringer},
  \& {Goldstein}}]{escoubet01}
{Escoubet} C.~P., {Fehringer} M., {Goldstein} M., 2001, \ang, 19, 1197

\bibitem[{{Gary} \& {Smith}(2009)}]{gary09}
{Gary} S.~P., {Smith} C.~W., 2009, \jgr, 114, A12105

\bibitem[{{Gosling}(2007)}]{gosling07a}
{Gosling} J.~T., 2007, \apjl, 671, L73

\bibitem[{{Gosling}(2012)}]{gosling12}
{Gosling} J.~T., 2012, \ssr, 172, 187

\bibitem[{{Gosling} \& {Phan}(2013)}]{gosling13}
{Gosling} J.~T., {Phan} T.~D., 2013, \apjl, 763, L39

\bibitem[{{Gosling} {et~al}\mbox{.}(2007){Gosling}, {Phan}, {Lin}, \&
  {Szabo}}]{gosling07b}
{Gosling} J.~T., {Phan} T.~D., {Lin} R.~P., {Szabo} A., 2007, \grl, 34, 15110

\bibitem[{{Gosling} \& {Szabo}(2008)}]{gosling08}
{Gosling} J.~T., {Szabo} A., 2008, \jgr, 113, 10103

\bibitem[{{Greco} {et~al}\mbox{.}(2008){Greco}, {Chuychai}, {Matthaeus},
  {Servidio}, \& {Dmitruk}}]{greco08a}
{Greco} A., {Chuychai} P., {Matthaeus} W.~H., {Servidio} S., {Dmitruk} P.,
  2008, \grl, 35, 19111

\bibitem[{{Greco} {et~al}\mbox{.}(2009){Greco}, {Matthaeus}, {Servidio},
  {Chuychai}, \& {Dmitruk}}]{greco09a}
{Greco} A., {Matthaeus} W.~H., {Servidio} S., {Chuychai} P., {Dmitruk} P.,
  2009, \apjl, 691, L111

\bibitem[{{Haynes}, {Burgess} \& {Camporeale}(2014){Haynes}, {Burgess}, \&
  {Camporeale}}]{haynes14}
{Haynes} C.~T., {Burgess} D., {Camporeale} E., 2014, \apj, 783, 38

\bibitem[{{Horbury} \& {Balogh}(1997)}]{horbury97a}
{Horbury} T.~S., {Balogh} A., 1997, \npg, 4, 185

\bibitem[{{Horbury} {et~al}\mbox{.}(2001){Horbury}, {Burgess}, {Fr{\"a}nz}, \&
  {Owen}}]{horbury01b}
{Horbury} T.~S., {Burgess} D., {Fr{\"a}nz} M., {Owen} C.~J., 2001, \grl, 28,
  677

\bibitem[{{Horbury}, {Wicks} \& {Chen}(2012){Horbury}, {Wicks}, \&
  {Chen}}]{horbury12}
{Horbury} T.~S., {Wicks} R.~T., {Chen} C.~H.~K., 2012, \ssr, 172, 325

\bibitem[{{Howes}(2015)}]{howes15}
{Howes} G.~G., 2015, \ptrsa, 373, 20140145

\bibitem[{{Howes} {et~al}\mbox{.}(2006){Howes}, {Cowley}, {Dorland}, {Hammett},
  {Quataert}, \& {Schekochihin}}]{howes06}
{Howes} G.~G., {Cowley} S.~C., {Dorland} W., {Hammett} G.~W., {Quataert} E.,
  {Schekochihin} A.~A., 2006, \apj, 651, 590

\bibitem[{{Howes}, {Klein} \& {TenBarge}(2014){Howes}, {Klein}, \&
  {TenBarge}}]{howes14b}
{Howes} G.~G., {Klein} K.~G., {TenBarge} J.~M., 2014, \apj, 789, 106

\bibitem[{{Johnstone} {et~al}\mbox{.}(1997){Johnstone}, {Alsop}, {Burge},
  {Carter}, {Coates}, {Coker}, {Fazakerley}, {Grande}, {Gowen}, {Gurgiolo},
  {Hancock}, {Narheim}, {Preece}, {Sheather}, {Winningham}, \&
  {Woodliffe}}]{johnstone97}
{Johnstone} A.~D. {et~al.}, 1997, \ssr, 79, 351

\bibitem[{{Kiyani} {et~al}\mbox{.}(2009){Kiyani}, {Chapman}, {Khotyaintsev},
  {Dunlop}, \& {Sahraoui}}]{kiyani09a}
{Kiyani} K.~H., {Chapman} S.~C., {Khotyaintsev} Y.~V., {Dunlop} M.~W.,
  {Sahraoui} F., 2009, \prl, 103, 075006

\bibitem[{{Klein}, {Howes} \& {TenBarge}(2014){Klein}, {Howes}, \&
  {TenBarge}}]{klein14b}
{Klein} K.~G., {Howes} G.~G., {TenBarge} J.~M., 2014, \apjl, 790, L20

\bibitem[{{Kunz} {et~al}\mbox{.}(2015){Kunz}, {Schekochihin}, {Chen}, {Abel},
  \& {Cowley}}]{kunz15}
{Kunz} M.~W., {Schekochihin} A.~A., {Chen} C.~H.~K., {Abel} I.~G., {Cowley}
  S.~C., 2015, \jplp, 81, 325810501

\bibitem[{{Landau} \& {Lifshitz}(1960)}]{landau60}
{Landau} L.~D., {Lifshitz} E.~M., 1960, {Electrodynamics of Continuous Media}.
  Pergamon Press

\bibitem[{{Leamon} {et~al}\mbox{.}(2000){Leamon}, {Matthaeus}, {Smith}, {Zank},
  {Mullan}, \& {Oughton}}]{leamon00}
{Leamon} R.~J., {Matthaeus} W.~H., {Smith} C.~W., {Zank} G.~P., {Mullan} D.~J.,
  {Oughton} S., 2000, \apj, 537, 1054

\bibitem[{{Leamon} {et~al}\mbox{.}(1998){Leamon}, {Smith}, {Ness}, {Matthaeus},
  \& {Wong}}]{leamon98a}
{Leamon} R.~J., {Smith} C.~W., {Ness} N.~F., {Matthaeus} W.~H., {Wong} H.~K.,
  1998, \jgr, 103, 4775

\bibitem[{{Loureiro}, {Schekochihin} \& {Zocco}(2013){Loureiro},
  {Schekochihin}, \& {Zocco}}]{loureiro13}
{Loureiro} N.~F., {Schekochihin} A.~A., {Zocco} A., 2013, \prl, 111, 025002

\bibitem[{{Malaspina} \& {Gosling}(2012)}]{malaspina12}
{Malaspina} D.~M., {Gosling} J.~T., 2012, \jgr, 117, 4109

\bibitem[{{Matthaeus} {et~al}\mbox{.}(2003){Matthaeus}, {Dmitruk}, {Oughton},
  \& {Mullan}}]{matthaeus03a}
{Matthaeus} W.~H., {Dmitruk} P., {Oughton} S., {Mullan} D., 2003, \aipcp, 679,
  427

\bibitem[{{McCracken} \& {Ness}(1966)}]{mccracken66}
{McCracken} K.~G., {Ness} N.~F., 1966, \jgr, 71, 3315

\bibitem[{{Meyrand} \& {Galtier}(2013)}]{meyrand13}
{Meyrand} R., {Galtier} S., 2013, \prl, 111, 264501

\bibitem[{{Neugebauer} {et~al}\mbox{.}(1984){Neugebauer}, {Clay}, {Goldstein},
  {Tsurutani}, \& {Zwickl}}]{neugebauer84}
{Neugebauer} M., {Clay} D.~R., {Goldstein} B.~E., {Tsurutani} B.~T., {Zwickl}
  R.~D., 1984, \jgr, 89, 5395

\bibitem[{{Ogilvie} {et~al}\mbox{.}(1995){Ogilvie}, {Chornay}, {Fritzenreiter},
  {Hunsaker}, {Keller}, {Lobell}, {Miller}, {Scudder}, {Sittler}, {Torbert},
  {Bodet}, {Needell}, {Lazarus}, {Steinberg}, {Tappan}, {Mavretic}, \&
  {Gergin}}]{ogilvie95}
{Ogilvie} K.~W. {et~al.}, 1995, \ssr, 71, 55

\bibitem[{{Osman} {et~al}\mbox{.}(2014){Osman}, {Matthaeus}, {Gosling},
  {Greco}, {Servidio}, {Hnat}, {Chapman}, \& {Phan}}]{osman14b}
{Osman} K.~T., {Matthaeus} W.~H., {Gosling} J.~T., {Greco} A., {Servidio} S.,
  {Hnat} B., {Chapman} S.~C., {Phan} T.~D., 2014, \prl, 112, 215002

\bibitem[{{Owens}, {Wicks} \& {Horbury}(2011){Owens}, {Wicks}, \&
  {Horbury}}]{owens11}
{Owens} M.~J., {Wicks} R.~T., {Horbury} T.~S., 2011, \solphys, 269, 411

\bibitem[{{Paschmann} {et~al}\mbox{.}(2013){Paschmann}, {Haaland}, {Sonnerup},
  \& {Knetter}}]{paschmann13}
{Paschmann} G., {Haaland} S., {Sonnerup} B., {Knetter} T., 2013, \ang, 31, 871

\bibitem[{{Perri} {et~al}\mbox{.}(2012){Perri}, {Goldstein}, {Dorelli}, \&
  {Sahraoui}}]{perri12a}
{Perri} S., {Goldstein} M.~L., {Dorelli} J.~C., {Sahraoui} F., 2012, \prl, 109,
  191101

\bibitem[{{Phan} {et~al}\mbox{.}(2010){Phan}, {Gosling}, {Paschmann}, {Pasma},
  {Drake}, {{\O}ieroset}, {Larson}, {Lin}, \& {Davis}}]{phan10}
{Phan} T.~D. {et~al.}, 2010, \apjl, 719, L199

\bibitem[{{Retin{\`o}} {et~al}\mbox{.}(2007){Retin{\`o}}, {Sundkvist},
  {Vaivads}, {Mozer}, {Andr{\'e}}, \& {Owen}}]{retino07}
{Retin{\`o}} A., {Sundkvist} D., {Vaivads} A., {Mozer} F., {Andr{\'e}} M.,
  {Owen} C.~J., 2007, \natphys, 3, 236

\bibitem[{{Russell}(1972)}]{russell72}
{Russell} C.~T., 1972, in NASA Special Pub. 308, Solar Wind, {Sonett} C.~P.,
  {Coleman} P.~J., {Wilcox} J.~M., eds., p. 365

\bibitem[{{Sahraoui} {et~al}\mbox{.}(2013){Sahraoui}, {Huang}, {Belmont},
  {Goldstein}, {R{\'e}tino}, {Robert}, \& {De Patoul}}]{sahraoui13a}
{Sahraoui} F., {Huang} S.~Y., {Belmont} G., {Goldstein} M.~L., {R{\'e}tino} A.,
  {Robert} P., {De Patoul} J., 2013, \apj, 777, 15

\bibitem[{{Schekochihin} {et~al}\mbox{.}(2009){Schekochihin}, {Cowley},
  {Dorland}, {Hammett}, {Howes}, {Quataert}, \& {Tatsuno}}]{schekochihin09}
{Schekochihin} A.~A., {Cowley} S.~C., {Dorland} W., {Hammett} G.~W., {Howes}
  G.~G., {Quataert} E., {Tatsuno} T., 2009, \apjs, 182, 310

\bibitem[{{Servidio} {et~al}\mbox{.}(2009){Servidio}, {Matthaeus}, {Shay},
  {Cassak}, \& {Dmitruk}}]{servidio09}
{Servidio} S., {Matthaeus} W.~H., {Shay} M.~A., {Cassak} P.~A., {Dmitruk} P.,
  2009, \prl, 102, 115003

\bibitem[{{Smith} {et~al}\mbox{.}(2006){Smith}, {Hamilton}, {Vasquez}, \&
  {Leamon}}]{smith06a}
{Smith} C.~W., {Hamilton} K., {Vasquez} B.~J., {Leamon} R.~J., 2006, \apjl,
  645, L85

\bibitem[{{Sundkvist} {et~al}\mbox{.}(2007){Sundkvist}, {Retin{\`o}},
  {Vaivads}, \& {Bale}}]{sundkvist07}
{Sundkvist} D., {Retin{\`o}} A., {Vaivads} A., {Bale} S.~D., 2007, \prl, 99,
  025004

\bibitem[{{Swisdak} {et~al}\mbox{.}(2010){Swisdak}, {Opher}, {Drake}, \&
  {Alouani Bibi}}]{swisdak10}
{Swisdak} M., {Opher} M., {Drake} J.~F., {Alouani Bibi} F., 2010, \apj, 710,
  1769

\bibitem[{{Swisdak} {et~al}\mbox{.}(2003){Swisdak}, {Rogers}, {Drake}, \&
  {Shay}}]{swisdak03}
{Swisdak} M., {Rogers} B.~N., {Drake} J.~F., {Shay} M.~A., 2003, \jgr, 108,
  1218

\bibitem[{{Taylor}(1938)}]{taylor38}
{Taylor} G.~I., 1938, \rslpsa, 164, 476

\bibitem[{{Telloni}, {Bruno} \& {Trenchi}(2015){Telloni}, {Bruno}, \&
  {Trenchi}}]{telloni15}
{Telloni} D., {Bruno} R., {Trenchi} L., 2015, \apj, 805, 46

\bibitem[{{TenBarge} \& {Howes}(2013)}]{tenbarge13a}
{TenBarge} J.~M., {Howes} G.~G., 2013, \apjl, 771, L27

\bibitem[{{Tsurutani} {et~al}\mbox{.}(1996){Tsurutani}, {Ho}, {Arballo},
  {Smith}, {Goldstein}, {Neugebauer}, {Balogh}, \& {Feldman}}]{tsurutani96}
{Tsurutani} B.~T., {Ho} C.~M., {Arballo} J.~K., {Smith} E.~L., {Goldstein}
  B.~E., {Neugebauer} M., {Balogh} A., {Feldman} W.~C., 1996, \jgr, 101, 11027

\bibitem[{{{\v S}afr{\'a}nkov{\'a}} {et~al}\mbox{.}(2013){{\v
  S}afr{\'a}nkov{\'a}}, {N{\v e}me{\v c}ek}, {P{\v r}ech}, \&
  {Zastenker}}]{safrankova13a}
{{\v S}afr{\'a}nkov{\'a}} J., {N{\v e}me{\v c}ek} Z., {P{\v r}ech} L.,
  {Zastenker} G.~N., 2013, \prl, 110, 025004

\bibitem[{{Vasquez} {et~al}\mbox{.}(2007){Vasquez}, {Abramenko}, {Haggerty}, \&
  {Smith}}]{vasquez07b}
{Vasquez} B.~J., {Abramenko} V.~I., {Haggerty} D.~K., {Smith} C.~W., 2007,
  \jgr, 112, A11102

\bibitem[{{Va{\v i}nshte{\v i}n}(1973)}]{vainshtein73}
{Va{\v i}nshte{\v i}n} S.~I., 1973, \spjetp, 37, 73

\bibitem[{{Wan} {et~al}\mbox{.}(2012){Wan}, {Matthaeus}, {Karimabadi},
  {Roytershteyn}, {Shay}, {Wu}, {Daughton}, {Loring}, \& {Chapman}}]{wan12a}
{Wan} M. {et~al.}, 2012, \prl, 109, 195001

\bibitem[{{Wu} {et~al}\mbox{.}(2013){Wu}, {Perri}, {Osman}, {Wan}, {Matthaeus},
  {Shay}, {Goldstein}, {Karimabadi}, \& {Chapman}}]{wu13}
{Wu} P. {et~al.}, 2013, \apjl, 763, L30

\bibitem[{{Xu} {et~al}\mbox{.}(2015){Xu}, {Wang}, {Wei}, {Feng}, {Deng}, {Ma},
  {Zhou}, {Pang}, \& {Wong}}]{xu15b}
{Xu} X. {et~al.}, 2015, \scirep, 5, 8080

\bibitem[{{Zhang} {et~al}\mbox{.}(2015){Zhang}, {He}, {Tu}, {Yang}, {Wang},
  {Marsch}, \& {Wang}}]{zhang15}
{Zhang} L., {He} J., {Tu} C., {Yang} L., {Wang} X., {Marsch} E., {Wang} L.,
  2015, \apjl, 804, L43

\bibitem[{{Zhdankin}, {Boldyrev} \& {Mason}(2012){Zhdankin}, {Boldyrev}, \&
  {Mason}}]{zhdankin12b}
{Zhdankin} V., {Boldyrev} S., {Mason} J., 2012, \apjl, 760, L22

\bibitem[{{Zhdankin} {et~al}\mbox{.}(2012){Zhdankin}, {Boldyrev}, {Mason}, \&
  {Perez}}]{zhdankin12a}
{Zhdankin} V., {Boldyrev} S., {Mason} J., {Perez} J.~C., 2012, \prl, 108,
  175004

\bibitem[{{Zhdankin} {et~al}\mbox{.}(2014){Zhdankin}, {Boldyrev}, {Perez}, \&
  {Tobias}}]{zhdankin14}
{Zhdankin} V., {Boldyrev} S., {Perez} J.~C., {Tobias} S.~M., 2014, \apj, 795,
  127

\end{thebibliography}

\end{document}